\documentclass[aps,twocolumn,showpacs,preprintnumbers]{revtex4}
\usepackage{amsmath}
\usepackage{epsfig}
\usepackage{graphicx}

\begin{document}

\title{Linear and quadratic temperature dependence \\ of electronic specific heat for cuprates}
\author{P. Salas, F.J. Sevilla and M.A. Sol\'{\i}s}
\affiliation{Instituto de F\'{\i}sica, Apartado postal 20-364, Universidad Nacional Aut\'onoma de M\'exico, \\ 01000 M\'exico, D.F., MEXICO}

\begin{abstract}
We model cuprate superconductors as an infinite layered lattice structure which 
contains a fluid of paired and unpaired fermions. Paired 
fermions, which are the superconducting carriers, are considered as noninteracting 
zero spin bosons with a linear energy-momentum dispersion relation, which coexist 
with the unpaired fermions in a series of almost two dimensional slabs stacked in 
their perpendicular direction.
The inter-slab penetrable planes are simulated by a Dirac comb potential in the 
direction in which the slabs are stacked, while paired and unpaired electrons 
(or holes) are free to move parallel to the planes. 
Paired fermions condense at a BEC critical temperature at which a jump 
in their specific heat is exhibited, whose values are assumed equal to the 
superconducting critical temperature and the specific heat jump 
experimentally reported for YBaCuO$_{7-x}$ to fix our model parameters: 
the plane impenetrability and the fraction of superconducting charge carrier. 
We  straightforwardly obtain, near and under the superconducting temperature $T_c$,
the linear ($\gamma_e T$) and the quadratic ($\alpha T^2$) electronic specific heat 
terms, with $\gamma_e$ and $\alpha$ of the order of the latest experimental values
reported. 
After calculating the lattice specific heat (phonons) $C_{l}$ from the phonon spectrum 
data obtained from inelastic neutron scattering experiments, and added to the electronic 
(paired plus unpaired) $C_{e}$ component, we 
qualitatively reproduce the total specific heat below $T_c$, whose curve lies close to the experimental one, reproducing its exact value at $T_c$. 
 \newline

\end{abstract}

\maketitle


\section{Introduction}

From the discovery of cuprate High Temperature Superconductors in 1986 (HTSC) 
\cite{Bednorz} many efforts have been made to explain the nature of their 
microscopic behavior as they are not completely described by the BCS theory \cite{BCS}. 
The HTSC cuprates are perhaps the most studied both experimentally and 
theoretically until FeAs showed up. Their main characteristics can be summarized as follows:
a small coherence length, usually of one or two nanometers; they have preferably
either tetragonal or orthorombic crystallographic structures; and the Cooper
pairs, responsible for the superconductivity, move in the copper oxide planes which 
resemble a quasi-2D layered system. They also modify their structure \cite{Poole2007} 
and their critical temperature $T_{c}$ by changing the oxygen concentration, 
being this feature responsible for achieving or not superconductivity for a
same compound with different oxygen concentrations. 

Particularly, the specific heat of the YBa$_{2}$Cu$_{3}$O$_{7-x}$ cuprates,
where $x$ represents the oxygen dopage made with holes, has been widely studied 
and we want to point out four key characteristics  
which have been observed. 
First of all, even though it is not easy to observe, below $T_c$ there is a linear 
term $\gamma_e T$ in the electronic specific heat, with $\gamma _{e}$ the 
electronic specific heat parameter or Sommerfeld constant - sometimes referred to as 
$\gamma(0)$-, whose latest reported values are between $2 - 3$ mJ/mol K$^2$ 
\cite{WangPhysRevB2001}. This term is currently believed to come from the normal 
state electronic specific heat (see Ref. \cite{Fisher2007} and references therein). 
In the second place, also at temperatures below $T_c$, there is an $\alpha T^2$
term in zero magnetic field \cite{Wright99,Moler97,WangPhysRevB2001} (initially 
denied in some reports \cite{Wright96}), which changes to a $H^{1/2}T$ component in the 
presence of an external magnetic field $H$, and is attributable to the superconducting 
part of the electronic specific heat. The reported values for this constant 
are of the order of tenths of mJ/mol K$^3$ \cite{WangPhysRevB2001}, but as it happens 
for the linear term coefficient, the obtained values depend strongly on the conditions 
of each experiment and on the theoretical method each author uses to subtract the 
other specific heat components. In the third place we have a ``jump'' (at
zero magnetic field) in the specific heat at $T_c$ \cite{Junod97} indicating a 
second order phase transition (which becomes a ``peak'' at finite magnetic field)
and is widely believed to be the result of the influence of the
superconducting electronic specific heat $C_{es}$. 
Finally, the total specific heat divided by temperature $C/T$  
shows an ``upturn'' (or ``fishtail'') for temperatures under 5 K, which several authors 
\cite{Fisher2007} claim has an intrinsic magnetic origin, present even 
at zero external magnetic field.  
This latter low temperature feature has been modeled as a contribution of five
terms: normal electronic $C_{en}$, superconducting electronic $C_{es}$,
lattice $C_{l}$, magnetic and hyperfine specific heats \cite{Fisher88,Fisher2007}, 
even though the dynamic mechanism beneath them is not completely known. 

Contrary to what one might think, to our knowledge, the lattice specific heat $C_{l}$ 
of a cuprate is not fully described by standard theories. This specific heat is 
generally considered as a contribution that doesn't change with the superconductivity 
onset and it has been observed to behave as a $T^3$ term below 5 K 
\cite{WangPhysRevB2001}. 
For the lattice specific heat a number of proposals have been made, such as: use 
of the Born-von K\'{a}rm\'{a}n formalism \cite{Kittel} 
applied to the multiatomic anisotropic lattice; fits of Debye and/or Einstein 
models \cite{Indios,Inderhees87,Junod89}; the use of different polynomials or power  
series on a number of different variables 
\cite{Inderhees92,Phillips1992,Gordon89,Bessergenev}; adaptation of models that have 
given successful results in other elements (such as $^4$He) 
\cite{Roulin96, Roulin98, Junod99}; the use of the lower Landau level (LLL) 
formalism \cite{Junod99}, to mention a few.

On the other hand, experimentalists have been using indirect methods 
for obtaining the electronic specific heat $C_{e}$ - although a
distinction between the normal and the superconducting part is sometimes 
not recognized - usually by separating  
the lattice specific heat from the total.
For example, Loram \textit{et al.} \cite{Loram93}, construct the lattice specific 
heat using a non-superconducting reference sample, using either a
small variation on the oxygen content or introducing another element; or Meingast 
\textit{et al.} \cite{Meingast09}, who construct their phonon density of states 
using a local-density approximation; or Bessergenev \textit{et al.} 
\cite{Bessergenev}, who develop a power series in terms of characteristic
temperatures which depend on phononic moments; or using an estimated phonon
spectrum based on known lattice vibration frequencies and inserting it in a set
of Einstein functions with characteristic temperatures \cite{Shaviv}. 
In all these studies, the authors subtract the ``lattice'' specific heat obtained 
from the total experimental specific heat of the
cuprate, and what is left is reported as the ``electronic specific heat'', 
which appears as a small contribution, restricted mainly to the height of the jump 
of only a few percent ($1 - 2 \%$), and unwillingly 
transferring the intrinsic uncertainties of their method to $C_{e}$. 
Based on our results, we propose that this manipulation should be reconsidered, since 
in this work we show that the electronic specific heat (normal and superconducting) 
contributes with a ($30 - 40$)$\%$ of the total.

For conventional superconductors $\gamma _{e}$ can be obtained from the linear 
term of the electronic specific heat in the normal state $C_{en}$, while the 
superconducting component $C_{es}$ has an exponential behavior at very low 
temperatures predicted by BCS. For HTSC this term has been confirmed to exist even 
in the superconducting state, however, it is very difficult to 
separate it from the other components due to thermal fluctuations  
and to the need of very high external magnetic fields to suppress 
the characteristic upturn \cite{WangPhysRevB2001,Fisher2007}. 
The importance of determining the value of $\gamma _{e}$ lies on its direct 
relation to the electronic density of states and on the belief that it gives 
information about the interaction electron-phonon \cite{Fisher2007}. In addition, 
there are models, such as Anderson's {\it Resonating Valence Bond} (RVB) 
\cite{Anderson}, that predict a linear term in the specific heat, 
but until now, the controversy over whether $\gamma _{e}$ comes from a
residual zero-field term, from the superconducting electronic specific heat 
\cite{Fisher2007} or from normal electrons that didn't participate in the
superconducting state (see, for example, \cite{UherCardwell} p.85) goes on. In 
this work we adopt this point of view and consider the linear part of the electronic 
specific heat, $\gamma _{e}T$, is a result of unpaired electrons inside a layered 
structure, as we will show. 

Even though conventional superconductors do not exhibit an $\alpha T^2$ electronic 
specific heat term, it has experimentally been confirmed that HTSC do 
\cite{WangPhysRevB2001}. Most of the reported values for this constant 
were obtained by first fixing $\gamma _{e}$ and then fitting curves with other 
parameters \cite{Wright99,Moler97}, while others \cite{Revaz98,WangPhysRevB2001} 
set an arrange of different external magnetic fields to cancel the interfering 
components. In the model presented in this paper, such a quadratic in temperature 
term in the electronic specific heat comes from paired electrons (superconducting 
state), and values of $\alpha$ are of the order of some ones reported 
experimentally \cite{Bessergenev}. 

At zero external magnetic field, the ``jump'' $\Delta C $ in the specific heat 
at the transition temperature has also been reported with a great variety of 
values depending on each experiment. Currently, this feature is attributed to the 
electronic specific heat, its magnitude is of $\Delta C \approx 5$ J/mol K according 
to several authors \cite{Gordon89,Inderhees92,Shaviv,Junod89}, and it has also been 
shown that for the same sample its jump diminishes as the magnitude of an external 
magnetic field is increased \cite{Junod97}. 
As experiments have become more accurate, the shape of the jump has become 
sharper \cite{WangPhysRevB2001}. However, a roundness of the curves at 
the jump with positive curvature is usually justified by the finiteness of the 
samples and by the presence of thermal fluctuations, which are 
important in HTSC \cite{Inderhees91}. 

In the framework of the most basic Boson-Fermion model 
\cite{Friedberg1989a,Friedberg1989b,Casas2001,Adhikari2000} of
superconductivity, we assume Cooper pairs are composite-spin-zero-bosons 
with either zero or nonzero moments of center of mass, coexisting with a 
fermion fluid formed by the unpaired electrons. These Cooper pairs are 
pre-formed at some temperature $T^{\ast} > T_c$ and can undergo 
a Bose-Einstein condensation (BEC) as temperature is lowered \cite{Casas2001,Eagles}. 
We are aware that the number of preformed pairs increases as the temperature is 
lowered until $T_c$, where its density is large enough to achieve coherence, 
independently of the mechanism by which the pairs are formed. Below $T_c$ we assume 
that the number of pairs remains constant. 

On the other hand, in order to include the effect of the layered structure of cuprates 
in the Boson-Fermion model, we previously calculated the BEC critical temperature and 
the thermodynamic properties for a system of non-interacting bosons immersed in a 
periodic multilayer array which represents the confinement agent \cite{PatyJLTP2010,
Paty2}. The multilayer array is simulated by an external Kronig-Penney (KP) potential at the 
delta limit case along the perpendicular direction to the CuO$_2$  planes, while the 
particles are allowed to move freely 
in the parallel directions with an energy-momentum dispersion relation where the linear term predominates, 
as has been shown in [\onlinecite{Adhikari2000}]. 

Our system model begins with $N$ electrons of mass $m_e$ interacting via 
a BCS type potential. There is a subgroup of electrons able to form pairs 
(Cooper pairs), since they are 
within a shell of width $2\hbar \omega_D$ around the Fermi energy $E_F$, 
where $\hbar \omega_D$ is the Debye energy, coexisting with a non-pairable 
group of electrons formed by those under and above the pairing shell, 
and are not eligible for pairing. 
From the first group, which we call {\it the pairable} electrons, we consider 
that only a fraction of them are {\it paired}, these are equal to a smaller fraction $f N/2$ which participate in the superconductivity; our assumption is based on 
the analysis of Uemura's plot (Fig. 2 of Ref \cite{Uemura2006}) that shows 
that critical temperatures for cuprates are in the empirical range of $T_c \approx 
(0.01 - 0.06) T_F$ \cite{AdhikariPhysC2000}. With all the stated above in mind, 
the $N$ electrons are grouped in three major components: paired electrons  (boson gas) formed by a fraction $f$ of half the total $N$  
electrons (inside the pairing shell); a fermion gas formed by the pairable but unpaired electrons (also inside the pairing shell); and the 
unpairable electrons (outside the pairing shell); 
plus a phonon gas due to the lattice. In Sec. \ref{GP} we obtain the grand potential 
from where it is possible to derive all the thermodynamic properties.
This model depends on three physical properties: the separation between planes $a$; 
the  impenetrability $P_{0}$ of the planes, which is responsible for the anisotropy 
observed in cuprates; and the density of superconducting carriers $f n/2$, with $n$ 
the fermionic number density. In 
section \ref{Crit Temp} we fix $a$ with the experimental values reported and 
calculate the critical temperature of the boson gas made of Cooper pairs as a 
function of $P_{0}$ and $f$.

In section \ref{SuperconductingCe} we obtain the {\it superconducting} electronic 
specific heat for the Cooper pairs fixing the unknown parameters $P_{0}$ and $f$ with 
the experimental $T_{cexp}$ and the magnitude of the ``jump'' in the electronic specific 
heat at $T_{c}$. As a consequence $C_{es}$ shows the observed $T^2$ behavior. Meanwhile, 
in section \ref{NormalCe} we derive the expressions for the {\it normal} electronic 
specific heat of the total unpaired $(1 - f) N$ electrons (fermions) in a periodic layered 
structure, which shows the expected linear dependence on $T$. In 
Sec. \ref{TotalCe} we add the two contributions and show that the total electronic 
specific heat at and under $T_c$ has the same behavior obtained by experiments. 
The electronic specific constants we obtain are $\gamma_e$ = 5.2 mJ/mol K$^2$ and 
$\alpha$ = 4.3 mJ/mol K$^3$, compared to 2.19  and 0.21 reported in 
\cite{WangPhysRevB2001}, and to 25.1 and 3.4 reported in \cite{Bessergenev}. 
In addition, since the lattice specific heat represents the main contribution of the 
total specific heat, in section \ref{Lattice} we calculate the specific heat for the 
phonons using two different approaches: Debye model and phonon density of states from 
inelastic neutron scattering (INS) experiments \cite{Renker88}. 
In section \ref{Sum}, we sum these three specific heat contributions and compare the 
result with experiments. We find an excellent qualitative agreement with the 
experimental $C$, where the contribution of the electronic specific heat is 
a significant part of the total. 
Finally, in Sec. \ref{conclusions} we present our conclusions.

\section{\label{The system}Cooper pairs in a layered structure}

We begin by taking a group of pairable electrons immersed 
in a periodic layered array along the $z$ direction and free to move in the 
other two directions with a linear 
dispersion relation. The wave vector for the center of mass of the pair 
(CMM) is given by $\mathbf{K} = (K_{x},K_{y},K_{z}) \equiv \mathbf{k_1} + 
\mathbf{k_2}$, while $\mathbf{k} \equiv 1/2(\mathbf{k_1} - \mathbf{k_2})$ is 
the relative momentum, where $\mathbf{k_1}$ and $\mathbf{k_2}$ are the wave 
vectors of each electron of the pair.
The solution for the Schr\"{o}dinger equation for the pairs may be separated
in the $x - y$ and $z$ directions, so the energy for each particle  is 
$\varepsilon_{K} = \varepsilon _{K_{x,y}} + \varepsilon _{K_{z}}$.
Here the total pair energy {\it in the plane} is $\varepsilon _{K_{x,y}} 
\equiv 2E_F - \Delta_K$, with $\Delta_K$ the binding energy of the pair for 
any temperature and any center of mass momentum. It has been
shown that when $\mathbf{K}$ is non zero, but small, one
can expand the binding energy from the Cooper equation in a series of powers 
\cite{Adhikari2000}, so the total energy in the plane is 
\begin{equation}
\varepsilon _{K_{x,y}} = \mathsf{e}_0 + C_{1}(K_{x}^{2}+K_{y}^{2})^{1/2}
+ O(K^2),  \label{Green0}
\end{equation}%
where $\mathsf{e}_0 \equiv 2E_F - \Delta_0$ is a constant, $C_{1}=(2/\pi )\hbar
\mathsf{v}_{F2D}$ is the linear term coefficient in 2D, $\mathsf{v}_{F2D}$ 
is the corresponding Fermi velocity, $\Delta_0 = 2 \hbar \omega_D \exp(-1/\lambda)$
is the energy gap for $\mathbf{K} = 0$ and weak coupling (corresponding to
the BCS theory), and $\lambda \equiv g(E_F)V$ the dimensionless coupling
constant in terms of the electronic density of states at the Fermi sea $g(E_F)$ 
and the non-local interaction $V$. 

Along the $z$-direction we use the Kronig-Penney \cite{KP} potential, where the 
energies are implicitly obtained, as a function of $a$, from the transcendental 
equation 
\begin{equation}
P_{0}(a/\lambda _{0})\sin (\alpha_{K_{z}} a)/\alpha_{K_{z}} a+
\cos (\alpha_{K_{z}} a)=\cos (K_{z}a),
\label{KPdelta}
\end{equation}
with 
$\alpha_{K_{z}} ^{2}\equiv
2m\varepsilon _{K_{z}}/\hbar ^{2}$, $m = 2 m_e$ the mass of the
composite-boson, and we have defined the dimensionless {\it plane impenetrability}  
$P_{0}\equiv P\lambda _{0}/a = m\Lambda\lambda _{0}/\hbar ^{2}$. 
The constant $\lambda _{0} \equiv h/\sqrt{2\pi mk_{B}T_{0}}$ is the de Broglie 
thermal wavelength of an ideal boson gas in an infinite box at the BEC 
critical temperature $T_{0} = 2\pi \hbar ^{2}n_{B}^{2/3}/mk_{B}
\zeta (3/2)^{2/3}\simeq 3.31\hbar ^{2}n_{B}^{2/3}/mk_{B}$, with 
$n_{B}\equiv N/(2 L^{3})$ the boson number density and $\Lambda$ is 
the strength of the KP delta potentials 
{$\sum_{n_z=-\infty }^{\infty } \Lambda \delta (z - n_za)$}. 

Note that when $P_{0}\rightarrow 0$ the energy goes to the 
free-particle energy $\varepsilon_{K_z}\rightarrow \hbar ^{2}K_{z}^{2}/2m$ in the  
$z$ direction. Also, when we have small energies, $\varepsilon _{K_{z}}<$ 
$\hbar ^{2}\pi^2/2ma^{2}$, we get the approximation  
\begin{equation}
\varepsilon _{K _{z}} \cong \varepsilon _{0}+\frac{\hbar ^{2}}{Ma^{2}}
(1-\cos K_{z}a),
\label{energiapeque}
\end{equation}
where $\varepsilon _{0} \equiv \hbar ^{2}{\alpha_{0}}^{2}/2m$ is the solution 
of Ec. (\ref{KPdelta}) when $K_{z}\rightarrow 0$ and $M$ is an effective mass 
\cite{Paty2,PatyJLTP2010}. Ec. (\ref{energiapeque}) is the most commonly 
used expression for quasi-bidimensional models of superconductors (see for example 
\cite{WK}), but it is a very limited model, since it involves calculations only 
over the first energy band and with $\varepsilon _{0} = 0$ the ground state energy. 
In this work we use the exact solution of the Eq. (\ref{KPdelta}).

\subsection{\label{GP}Grand potential}

To calculate the thermodynamic properties we begin with the grand potential 
$\Omega (T,L^3,\mu )$, which for a boson gas with $N_B$ particles contained in a
volume $V\equiv L^{3}$ \cite{Path} is  
\begin{gather}
\Omega (T,L^3,\mu )=U - TS-\mu N_{B}  
=\Omega _{0} + k_{B}T\sum_{\mathbf{K}{\neq 0}}\ln \bigl\{1-\notag \\ 
\exp [-\beta 
(\mathsf{e}_0 + C_{1}(K_{x}^{2}+K_{y}^{2})^{1/2}+ 
\varepsilon _{K_{z}}- \mu )] \bigr\},  
\label{omega}
\end{gather}
where $U$ is the internal energy, $S$ the entropy, $\mu $ the chemical
potential, $\beta \equiv 1/k_{B}T$, and the first term in the rhs
corresponds to the $\mathbf{K}={0}$ ground state energy contribution 
$\Omega _{0}$ = $k_{B}T\ln \{1-\exp [-\beta (\varepsilon _0 + 
\mathsf{e}_0-\mu )]\} $.

Using that $\ln(1-x)= -\sum_{l=1}^{\infty} x^{l}/l$
in Ec. (\ref{omega}), and after some algebra we have  
\begin{gather}
\Omega (T,L^3,\mu )= \Omega _{0}  
- k_{B}T \sum_{l=1}^{\infty} \sum_{\mathbf{K}{\neq 0}}
\frac{\exp \bigl\{\beta (\mu - \mathsf{e}_0)l\bigr\}}{l} \notag \\
\times\exp \bigl\{-\beta (C_{1}(K_{x}^{2}+K_{y}^{2})^{1/2}+ 
\varepsilon _{K_{z}})l\bigr\}.  \label{Omegalin3}
\end{gather}

By substituting sums by integrals in the thermodynamic limit, and doing the integrals 
over $x,y$ one gets
\begin{gather}
\Omega \left( T,L^3,\mu \right)=k_{B}T\ln \bigl\{1- \exp[-\beta(\varepsilon _{0} 
+ \mathsf{e}_{0} -\mu)\bigr\}- \notag  \\
\frac{1}{\beta ^{3}}\frac{L^{3}}{\left( 2\pi \right) ^{2}}
\frac{\Gamma(2)}{C_{1}^{2}} {\int_{-\infty }^{\infty }dK_{z}} \ 
\mathsf{g}_{3}\bigl\{\exp[-\beta (\varepsilon_{K_{z}} + \mathsf{e}_{0} - 
\mu )]\bigr\}.  
\label{TGP1}
\end{gather}
where we have used the Bose functions $\mathsf{g}_{\sigma }(t) \equiv 
\sum_{l=1}^{\infty}
(t)^{\mathit{l}}/\mathit{l}^{\sigma}$. From (\ref{TGP1}) we can find
the thermodynamic properties for a monoatomic Bose gas \cite{PatyJLTP2010,Paty2}.

\subsection{ \label{Crit Temp}Critical temperature}

To obtain the critical temperature of the boson gas we use the expression
for the bosonic particle number, which we calculate summing the number of 
particles in each energy state, therefore 
\begin{gather}
N_B=\frac{1}{\exp \bigl\{\beta (\varepsilon _{0}+\mathsf{e}_{0}-\mu )
\bigr\}-1}+ \notag \\
\frac{L^{3}}{\left( 2\pi\right) ^{2}}\frac{\Gamma (2)}{C_{1}^{2}}\frac{1}
{\beta ^{2}}{\int_{-\infty}^{\infty }dK_{z}} \mathsf{g}_{2}\bigl\{ \exp 
[-\beta (\varepsilon _{K_{z}} + \mathsf{e}_{0}-\mu )]\bigr\}, \label{numlin2}
\end{gather}%
where the first term of the rhs corresponds to the number of particles in
the condensate $N_{0}(T)$ and the second term to the ones in the excited 
state $N_{e}(T)$.

We must notice here that from the relation for the Fermi energy \cite{Path}, 
$E_{F} = {\hbar^2}(3\pi^2)^{2/3} n_s^{2/3}/2m^{\ast}$, with $E_{F}$ the Fermi 
energy which corresponds to a Fermi temperature of $T_F$ = 2290 K for the 
cuprate and $m^{\ast} = 2 m_e$ the effective mass of the 
carriers, one gets that the density number of carriers is $n_{s} = 
1.128\times 10^{27}/$m$^{3}$. 
On the other hand, from the relation $T_{0}/T_F = 0.218$ \cite{Sevilla} one 
gets $T_{0} = 499.2$ K, when all the fermions in the cuprate are paired. Using 
this boson gas temperature and from the definition of $T_0$, one gets 
$n_{B}=1.994\times 10^{26}/$m$^{3}$, the boson density number of an ideal gas, 
whose value is an order of magnitude smaller than $n_{s}$. 
However, as we previously stated, 
analyzing the data in Fig. 2 of Ref \cite{Uemura2006}, and localizing the diagonal 
lines labeled as $T = T_F$ and $T = T_0$ (identified as $T_B$),   
one would expect that the actual quantity of superconducting carriers $n_b$ 
for the cuprate materials would be at least two orders of magnitude smaller than 
the $n_s$ given above. Therefore, we assume that only a 
fraction $f$ of the maximum possible value $n_{B}$ is participating in the boson 
gas responsible for the superconductivity, so $n_{b} = fn_{B}$, and we 
expect this factor to lie in the interval $f \in [0.01, 0.14]$, according to what 
is explained in Ref. \cite{AdhikariPhysC2000}. 

The BEC temperature of this boson gas in the thermodynamic limit is 
\begin{equation}
T_{0 f}=\frac{2\pi \hbar ^{2}n_{b}^{2/3}}{mk_{B}\zeta (3/2)^{2/3}}
 =T_{0}f^{2/3}.  \label{T0Bf}
\end{equation}
For $f = 1$ we recover the case where all pairable 
fermions participate in the boson gas. The corresponding thermal
wavelenght is $\lambda_{0f}=h/\sqrt{2\pi m k_{B}T_{0f}}= \lambda
_{0}/f^{1/3} $. The quotient of the fraction of an ideal 
gas BEC temperature over the Fermi temperature of the whole is \cite{Sevilla} 
\begin{equation}
\frac{T_{0f}}{T_{F}}=\frac{2\pi f^{2/3}}{(6\pi ^{2})^{2/3}\zeta (3/2)^{2/3}}
=(0.218)f^{2/3}.  \label{T0fsTF}
\end{equation}

Now we use the relation for the 3D Fermi energy $E_{F3{\text D}}=[(3\pi
^{2})^{2/3}/2\pi ]E_{F2{\text D}}$ for a 3D system in terms of the Fermi 
energy for a 2D system $E_{F2{\text D}}=\frac{1}{2}m_{e}\mathsf{v}_{F2D}^{2}$. 
The constant $C_{1}^{2}=(4/\pi ^{2})\hbar ^{2}\mathsf{v}_{F2D}^{2}=[64/\pi 
(3\pi^{2})^{2/3}]\gamma a^{2}E_{F3D}k_BT_0$, where $\gamma \equiv 
\hbar ^{2}/2ma^{2}k_{B}T_{0}$ is a dimensionless constant. Introducing $f$ 
in (\ref{numlin2}), dividing by $N$ and taking $T=T_{c}$, so the chemical 
potential $\mu_{0} = 
\varepsilon_0 + \mathsf{e}_{0}$ and $N_{0}(T_{c})\simeq 0$, we have 
\begin{gather}
1=\frac{2}{f}\frac{\Gamma (2) 3\pi
(3\pi ^{2})^{2/3}}{128} \left(\frac{0.436\gamma T_0}{T_F} \right)^{1/2} 
\frac{1}{(k_{B}T_{F}\beta_c)^{2}}  \notag  \\  
\times \int_{0}^{\infty}adK_{z}
\mathsf{g}_{2}\bigl\{\exp [-k_{B}T_{F}\beta_c\gamma (T_0/T_F)
(\bar{\varepsilon}_{K_z}- \bar{\varepsilon}_{0})]\bigr\}.  \label{TcLin}
\end{gather}
where $\bar{\varepsilon}_{K_z} \equiv  \varepsilon_{K_z}/(\hbar^2/2ma^2)$, and 
$\bar{\varepsilon}_{0} \equiv \varepsilon_0/(\hbar^2/2ma^2)$ are dimensionless 
energies obtained numerically from (\ref{KPdelta}) for each band. 

We split the infinite integral in (\ref{TcLin}) as a sum of integrals over 
the allowed energy bands, fold every band over the first half Brillouin zone 
(from $0$ to $\pi$) and cut the sum at the $J$-th band once convergence has 
been achieved. Finally, using the numerical value $\Gamma (2)3\pi
(3\pi ^{2})^{2/3}(0.436 T_0/T_F)^{1/2}/(128)=0.46532$ we arrive to 
\begin{gather}
1=\frac{2}{f}0.46532 \gamma ^{1/2} \frac{1}{%
(k_{B}T_{F}\beta _{c})^{2}}  \notag \\
\times \sum_{j=1}^{J}\int_{0}^{\pi }adK_{z}
\mathsf{g}_{2}\bigl\{\exp [-k_{B}T_{F}\beta_c\gamma (T_0/T_F)(\bar{\varepsilon}_
{K_z}- \bar{\varepsilon}_{0})]\bigr\},
\end{gather}%
which must be solved numerically. From now on we will be using this 
allowed-band-splitting method for evaluating the infinite integrals.
\begin{figure}[tbh]
\begin{center}
\centerline{\epsfig{file=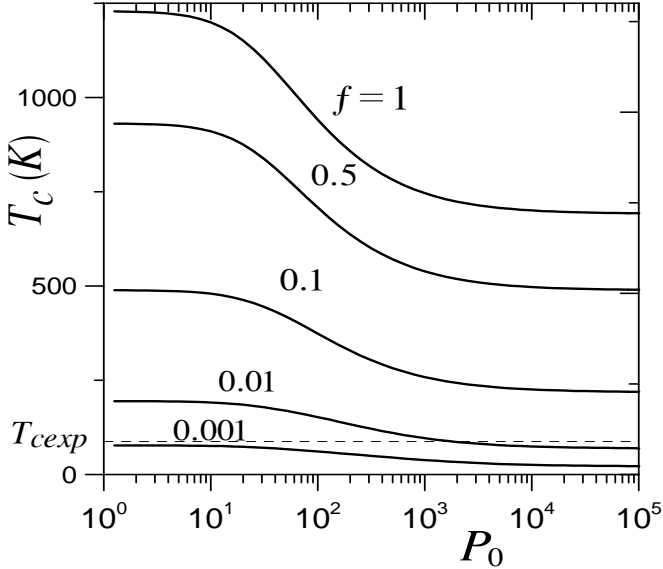,height=3.in,width=3.50in}}
\end{center}
\par
\vspace{-1.0cm}
\caption{Critical temperature as a function of $P_{0}$ for different values
of $f$. Dashed line is the experimental $T_{c} = 92.2$ K for the cuprate.}
\label{fig:Tclineal}
\end{figure}

The experimental parameters for the cuprate YBa$_{2}$Cu$_{3}$O$_{6.92}$ 
that we use are: the critical temperature $T_{c exp}$ = 92.2 K \cite{Bessergenev}; 
the Fermi temperature $T_{F}=2290$ K \cite{Poole95};  $\Delta_0 = 20$ meV 
\cite{WangPhysRevB2001}; the parameter $a=5.85$ \AA\ corresponding to $a=c/2$ 
($c=11.69$ \AA\ the crystallographic constant), the distance between a copper 
oxide plane located at the extreme of the unitary cell and the Ytrium atom at 
the center. This is based on the knowledge that superconductivity occurs in the 
close  vicinity parallel to the CuO$_{2}$ planes, which are two per unit cell. 
We can also calculate the thermal 
wavelength $\lambda _{0}=23.575$ \AA \thinspace, and  the parameters 
$a/\lambda _{0}=0.246$ and $\gamma =1.317$. Finally, we obtain the 
magnitud of the jump $\Delta C/T_c$ $\simeq$ 20 mJ/mol K$^2$ from the data 
published by \cite{Bessergenev}. It is well known that in cuprates there is an
optimum oxygen dopage for which $T_{c}$ is higher \cite{Roulin98}, but we
choose the dopage for which we find more accurate experimental data.

In Fig. \ref{fig:Tclineal} we show the critical temperature as a function of
the parameter $P_{0}$ for five values of $f$. The dashed line represents the
experimental critical temperature for YBa$_{2}$Cu$_{3}$O$_{6.92}$. 
As we can see from this figure, there is only a narrow interval
of values of $f \in [0.001,0.02]$, that suits the experimental condition  
$T_{c exp}$ = 92.2 K, consistent with what we expected, which in turn determines 
a set of values of $P_0$. This 
means that only a small percentage of the initially pairable fermions form pairs, 
as we argued above. To determine exactly both values, we choose a second feature 
of the electronic specific heat: the magnitude of the jump. 

\subsection{\label{Int Ener}Internal Energy}

We can derive the internal energy of the pair's gas as
\begin{gather}
U(T,V) =\frac{(\varepsilon_0 + \mathsf{e}_{0})}
{\exp \bigl\{\beta (\varepsilon_0 + \mathsf{e}_{0}-\mu)\bigr\}-1} 
+\frac{L^{3}}{\left( 2\pi \right) ^{2}}\frac{\Gamma (2)}{C_{1}^{2}}
\frac{1}{\beta ^{2}} \times  \notag \\
{\int_{-\infty}^{\infty }dK_{z}}( \mathsf{e}_{0}+
\varepsilon _{K_{z}}) \mathsf{g}_{2}\bigl\{ \exp[-\beta (\varepsilon _{K_{z}}+ 
\mathsf{e}_{0}-\mu )]\bigr\}  \notag \\
+2\frac{L^{3}}{\left( 2\pi \right) ^{2}}\frac{\Gamma (2)}{C_{1}^{2}}
\frac{2}{\beta ^{3}}{\int_{-\infty}^{\infty }dK_{z}} \mathsf{g}_{3}\bigl\{ \exp 
[-\beta (\varepsilon_{K_{z}}+\mathsf{e}_{0}-\mu )]\bigr\},  \label{Ulin}
\end{gather}%
where the first term corresponds to the particles in the ground state, 
$(\varepsilon_0 + \mathsf{e}_{0})N_0$. From the previous equation we subtract 
the ground state energy times the total number of pairs 
given by the number equation (\ref{numlin2}), then multiply the result by $f$
and divide it by $N_Bk_BT$, so we have
\begin{gather}
\frac{(U-(\varepsilon _{0} + \mathsf{e}_{0})N_B)}{N_Bk_{B}T} = 
\frac{2}{f} 0.46532 \left( \frac{2 T_{0}}{T_{F}}\right)
\gamma^{3/2}\times \notag \\
 \frac{1}{k_{B}T_{F}\beta} 
 {\int_{-\infty}^{\infty }adK_{z}}
(\bar{\varepsilon}_{K_{z}}-\bar{\varepsilon}_{0})\mathsf{g}_{2}\bigl\{\exp [-k_{B}T_{F}
\beta\gamma (T_{0}/T_{F}) \times \notag \\
(\bar{\varepsilon}_{K_{z}}+\bar{\mathsf{e}}_{0}-\bar{\mu})]
\bigr\} 
+ \frac{4}{f}0.46532\gamma ^{1/2}\frac{1}
{{(k_{B}T_{F}\beta)}^{2}}\notag \\
 \times {\int_{-\infty}^{\infty }}adK_{z} \mathsf{g}_{3}\{\exp [-k_{B}T_{F}\beta\gamma 
(T_{0}/T_{F})(\bar{\varepsilon}_{K_{z} }+\bar{\mathsf{e}}_0 -\bar{\mu})]\}. 
\label{Umenose0X}
\end{gather}
In order to compute the internal energy and the specific heat it is necessary to
get numerically the chemical potential and its derivative with respect to $T$. 
From the number equation (\ref{numlin2})
and by making the considerations that $\mu =\mu _{0}$ if $T<T_{c}$, and 
$N_{0}/N_{b}\sim 0$ for $T>T_{c}$, $\mu$ is obtained from 
\begin{gather}
1=\frac{2}{f}0.46532 \gamma ^{1/2} 
\frac{1}{(k_{B}T_{F}\beta)^{2}}  
 \int_{-\infty}^{\infty }adK_{z} \times \notag \\
\mathsf{g}_{2} \bigl\{\exp [-(k_{B}T_{F}\beta)
\gamma (T_{0}/T_{F})
(\bar{\varepsilon}_{k_{z}}+ \bar{\mathsf{e}}_{0}-\bar{\mu} )]\bigr\}, 
\label{mulin2}
\end{gather}
and its derivative is given by 
\begin{gather}
T\frac{d\bar{\mu}}{dT}= 
\biggl[2k_{B}T_{F}\beta-\frac{2}{f}0.46532 \gamma^{3/2}
\left( \frac{ T_{0}}{T_{F}}\right){\int_{-\infty}^{\infty }}
adK_{z}  \times \nonumber \\
\ln\{1-\exp [-k_{B}T_{F}\beta\gamma (T_{0}/T_{F})
(\bar{\varepsilon}_{K_{z}}-
\bar{\mu} +\bar{\mathsf{e}}_{0})]\} \times \nonumber \\
(\bar{\varepsilon}_{K_{z}}-\bar{\mu} +
\bar{\mathsf{e}}_{0})\biggr]/
\biggl[\frac{2}{f}0.46532 \gamma^{3/2}\left( \frac{2 T_{0}}{T_{F}}\right)
{\int_{-\infty}^{\infty }}adK_{z} \times \nonumber \\
\ln\{1-\exp [-k_{B}T_{F}\beta\gamma (T_{0}/T_{F})
(\bar{\varepsilon}_{K_{z}}-
\bar{\mu} + \bar{\mathsf{e}}_{0})]\} \biggr].  \label{Tdmulinf}
\end{gather}

\section{Electronic specific heat}

We consider that the electronic specific heat $C_{e}$ of the cuprate is 
formed by the specific heat of the gas of Cooper-pairs, plus the specific 
heat of the gas of all the remaining electrons that didn't form pairs.

Like most authors do, we assume that the electronic specific heat is
responsible for the jump, which in turn provides the information about the 
phase transition.

In this model approach, we also consider that the specific 
heat at constant volume $C_{V}$ is the same as the specific 
heat at constant pressure $C_{p}$ at least from $T = 0$ to $200$ K, as has been 
established for cuprates by several authors \cite{Meingast91,Nagel,Jorgensen,
Meingast09}. So from now on we will drop any subscript on this matter.

\subsection{\label{SuperconductingCe}Superconducting electronic specific heat}

We are able to get the superconducting electronic specific heat $C_{es}$ by 
introducing the internal energy (\ref{Ulin}) in 
$C_{es}=\left[ {\frac{\partial }{\partial T}}U(T,L^3)\right] _{N,L^3}$, and  
taking the fraction $f$  
\begin{gather}
\frac{C_{es}}{N_bk_{B}} = \frac{2}{f} 0.46532 \left( \frac{2 T_{0}}{T_{F}}\right)
\gamma^{3/2}\frac{1}{(k_{B}T_{F}\beta)} \notag  \\
\times \int_{-\infty}^{\infty } adK_{z} \ 
\mathsf{g}_{2}\{\exp [-k_{B}T_{F}\beta \gamma (T_{0}/T_{F})(\bar{\varepsilon}_{K_{z}} 
-\bar{\mathsf{e}}_0-\bar{\mu})]\} \notag \\
\times[2\bar{\varepsilon}_{K_{z}}-\bar{\varepsilon}_0+\bar{\mathsf{e}}_0-
\bar{\mu}+T\frac{d\bar{\mu}}{dT}]  \notag  \\
-\frac{1}{2f}0.46532\left( \frac{2 T_{0}}{T_{F}}\right)^2 \gamma^{5/2} 
\int_{-\infty}^{\infty }adK_{z}(\bar{\varepsilon}_{K_{z}}-
\bar{\varepsilon}_0) \notag  \\
\times \ln \{1-\exp [-k_{B}T_{F}\beta\gamma (T_{0}/T_{F})
(\bar{\varepsilon}_{K_{z}}+\bar{\mathsf{e}}_0 -\bar{\mu})]\}  \notag \\
\times [\bar{\varepsilon}_{K_{z}}+\bar{\mathsf{e}}_0-\bar{\mu}
+T\frac{d\bar{\mu}}{dT}] 
+\frac{12}{f}0.46532\gamma ^{1/2}\frac{1}
{{(k_{B}T_{F}\beta)}^{2}}  \notag  \\
\times \int_{-\infty}^{\infty }adK_{z} \mathsf{g}_{3}\{\exp [-k_{B}T_{F}\beta\gamma
(T_{0}/T_{F})(\bar{\varepsilon}_{K_{z}}+\bar{\mathsf{e}}_0-\bar{\mu})]\}.  
\label{Cvlinf}
\end{gather}

\begin{figure}[htb]
\begin{center}
\centerline{\epsfig{file=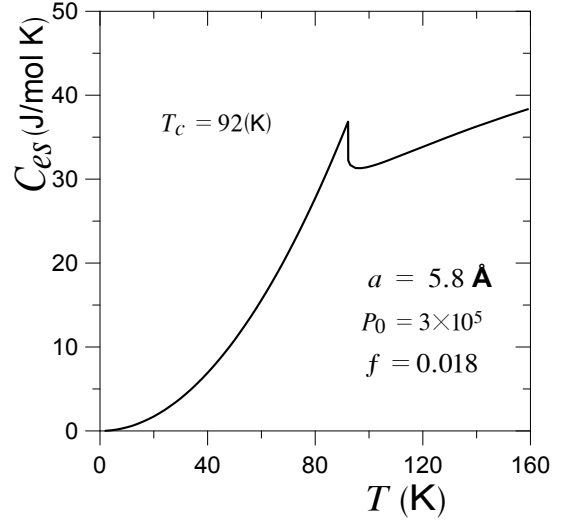,height=2.8in,width=3.0in}}
\end{center}
\vspace{-1.0cm}
\caption{Electronic specific heat as a function of $T$ for the Cooper 
pairs gas for YBa$_{2}$Cu$_{3}$O$_{6.92}$ using Ec. (\ref{Cvlinf}), using 
$f = 0.018$ and $P_0 = 3 \times 10^5$.}
\label{fig:CvBosonesLin}
\end{figure}

\begin{figure}[htb]
\begin{center}
\centerline{\epsfig{file=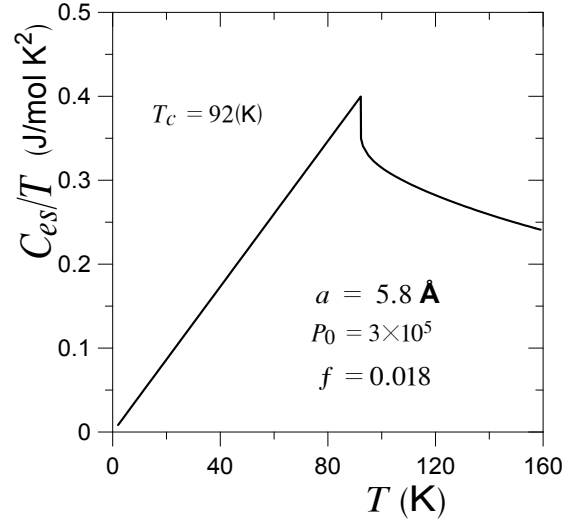,height=2.8in,width=3.0in}}
\end{center}
\vspace{-1.0cm}
\caption{Electronic specific heat over temperature as a function of $T$ 
for the Cooper pairs gas for YBa$_{2}$Cu$_{3}$O$_{6.92}$.}
\label{fig:CvsTBosonesLin}
\end{figure}

In Figs. \ref{fig:CvBosonesLin} and \ref{fig:CvsTBosonesLin} we present the
superconducting electronic specific heat $C_{es}$ and $C_{es}/T$ as function 
of temperature for the gas of Cooper pairs. 
The height of the jump is reproduced by taking the values $f = 0.018$ (which lies 
inside the interval [0.01, 0.14] obtained in Sec. \ref{Crit Temp}) and 
$P_0 = 3 \times 10^5$ for the dimensionless parameters of our model.


The quadratic behavior of $C_{es}$ with temperature is clear from the 
curve presented in Fig \ref{fig:CvsTBosonesLin}, so we can extract the slope 
$\alpha = 4.3$ mJ/mol K$^3$. It is also important to point out that 
the percentage of paired electrons that participate in superconductivity 
is among 1 and 3$\%$, consistent with the report in \cite{WangPhysRevB2001} 
and in contrast to the 90 to 95$\%$ reported in \cite{Shaviv}.

\subsection{\label{NormalCe}Normal electronic specific heat}

We shall consider now the gas of fermions made of the electrons with 
mass $m_e$ in the spherical shell who didn't pair to form Cooper pairs plus 
the unpairable electrons, constituting $(1 - f)$ of the total $N$ electrons. 
The grand potential for an ideal fermi gas comes from \cite{Path} 
\begin{equation}
\Omega (T,L^{3},\mu_{fer} )=-k_{B}T\sum_{\mathbf{k}{= 0}}\ln \bigl\{1+\exp 
[-\beta(\varepsilon _{\mathbf{k}}-\mu_{fer} )]\bigr\}.  \label{omegaFer}
\end{equation}
where $\mu_{fer}$ is the chemical potential of the electron gas and 
$\varepsilon _{\mathbf{k}} = \hbar^2 k_x^2/2m_e + \hbar^2 k_y^2/2m_e + 
\varepsilon_{k_z}$ is the energy of each electron free in the $x-y$ directions 
and constrained by permeable planes in $z$ direction. As we did in the 
case of the boson gas, the $z$-component energy must come from the KP 
Eq. (\ref{KPdelta}) taking the momentum of the electron $k_z$ and the 
corresponding impenetrability $P_{0F}$ = $P_{0}/2$. 
Replacing sums for integrals in the thermodynamic limit, and doing the 
integrals over $k_{x}$, $k_{y}$ we have 
\begin{gather}
\Omega \left( T,L^3,\mu _{fer}\right) =-2\frac{L^{3}}{\left( 2\pi \right) ^{2}}
\frac{m_{e}}{\hbar ^{2}}\frac{1}{\beta ^{2}}  \notag \\
\times {\int_{-\infty }^{\infty }dk_{z}}\mathsf{f}_{2}(\exp [-\beta
(\varepsilon _{k_{z}}-\mu _{fer})]),  \label{TGPferm}
\end{gather}
where we have made use of the Fermi-Dirac functions $\mathsf{f}_{\sigma }(t)
\equiv \sum_{l=1}^{\infty }(-1)^{l-1}t^{\mathit{l}}/
\mathit{l}^{\sigma }$ \cite{Path}.

In order to obtain the electronic specific heat for the fermions $C_{en}$ (sometimes 
referred to as $C_{lin}$ \cite{WangPhysRevB2001}), first 
we deduce the internal energy of the fermi gas and then the specific heat, so 
doing the usual algebra we have 
\begin{gather}
\frac{C_{en}}{N_{fer}k_{B}} =4\frac{1}{(1-f)}\frac{L^{3}}{N \left( 2\pi 
\right) ^{2}}\frac{m_e}{\hbar ^{2}}\frac{1}{\beta } \notag  \\
\times {\int_{-\infty }^{\infty }dk_{z}}\mathsf{f}_{2}\{\exp[-\beta 
(\varepsilon _{k_{z}}-\mu_{fer} )]\} + \nonumber \\
2\frac{1}{(1-f)}
\frac{L^{3}}{N \left( 2\pi \right) ^{2}}\frac{m_e}{\hbar ^{2}} 
\int_{-\infty }^{\infty }dk_{z}\ln \{1+ \notag  \\
\exp [-\beta (\varepsilon
_{k_{z}}-\mu_{fer} )]\} 
 \{2\varepsilon _{k_{z}}-\mu_{fer} +T\frac{\partial 
\mu_{fer}}{\partial T}\}  \notag \\
+2\frac{1}{(1-f)}\frac{L^{3}}{N \left( 2\pi \right) ^{2}}\frac{m_e}
{\hbar ^{2}}\beta \notag  \\
\times {\int_{-\infty }^{\infty }dk_{z}}\frac{\varepsilon _{k_{z}}
\{\varepsilon_{k_{z}}-\mu_{fer} +T\frac{\partial \mu_{fer} }{\partial T}\}}
{\exp [\beta (\varepsilon_{k_{z}}-\mu_{fer} )]+1}, 
\end{gather}
where $N_{fer}$ is the number of non-paired fermions.

The chemical potential is calculated from the corresponding number equation as
\begin{gather}
1=\frac{1}{(1-f)}\frac{L^{3}}{N }\frac{2}{\left( 2\pi \right) ^{2}}
\frac{m_{e}}{\hbar ^{2}}\frac{1}{\beta }  \notag \\
\times {\int_{-\infty }^{\infty }dk_{z}}\ln \{1+\exp [-\beta (\varepsilon
_{k_{z}}-\mu _{fer})]\}
\end{gather}
from where $\mu _{fer}$ and its derivative are extracted using numerical methods.

\begin{figure}[tbh]
\begin{center}
\centerline{\epsfig{file=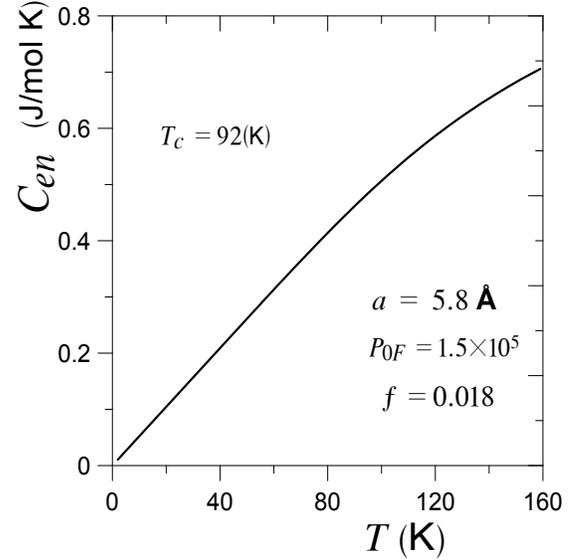,height=3.0in,width=3.0in}}
\end{center}
\par
\vspace{-1.0cm}
\caption{Specific heat as a function of $T$ for the unpaired electrons, 
using $f = 0.018$ and $P_{0F} = 1.5 \times 10^5$.}
\label{fig:Cvferm}
\end{figure}

In Fig. \ref{fig:Cvferm} we present the behavior of the normal electronic
specific heat for the unpaired electrons taking the values of 
($f,P_{0}$) obtained in the preceding section. Notice that $C_{en}$ is 
linear in $T$ until well above $T_{c}$, as expected, and its coeficient is 
$\gamma_e$ = 5.2 mJ/mol K$^{2}$. 

Clearly, there is a \textit{non-zero} value for $\gamma _{e}$, as stated 
in some reports \cite{Liang} for oxygen contents $x>0.6$,  and it
corresponds to the contribution of the unpaired electron, as suggested by 
Fisher, \textit{et al.} in \cite{Fisher2007}. 

The analysis of the thermodynamic properties of a gas of fermions in a 
layered structure will be made elsewhere.

\subsection{\label{TotalCe}Total electronic specific heat}

\begin{figure}[tbh]
\begin{center}
\centerline{%
\epsfig{file=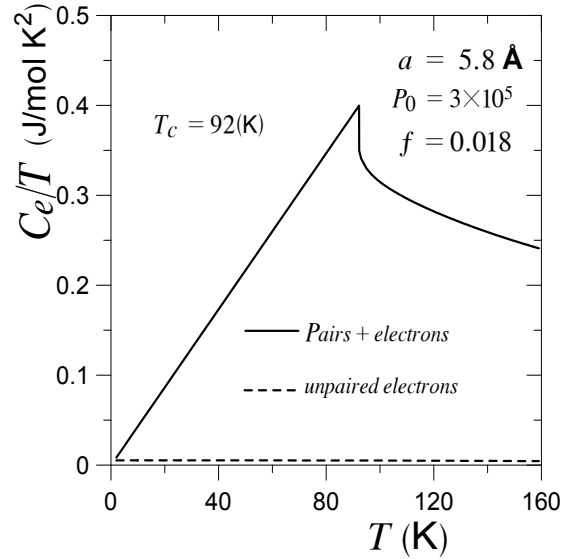,height=3.0in,width=3.0in}}
\end{center}
\par
\vspace{-1.0cm}
\caption{Specific heat over temperature $T$ for Cooper pairs plus unpaired
electrons, using $f = 0.018$ and $P_0 = 3 \times 10^5$. The dashed line is 
the unpaired electrons contribution.}
\label{fig:CelectronicosT}
\end{figure}

The total electronic specific heat is the sum of the previous contributions:
the Cooper pairs specific heat and the unpaired electrons specific heat,
$C_{e}=C_{es}+C_{en}$. We find that the contribution of the
normal state electronic specific heat is  two orders of magnitude smaller than 
the superconducting counterpart (Cooper pairs), so in Fig. \ref{fig:CelectronicosT} 
we present the curve $C_{e}/T$ {\it vs} $T$, where the difference can be better 
appreciated, the dashed line being the normal electronic specific heat. 

There are some considerations we would like to do at this point, based in 
our results.
First, we do not reproduce  
the upturn for the $T < 5$  K temperature region since the internal magnetic components and the 
influence of external magnetic fields are not considered, as expected. 
Second, we have the coexistence of the two gases at least for temperatures 
under $T_c$. Above $T_c$ paired fermions decouple in an unknown way, so we 
do not include the decoupling mechanism in our analysis. As a consequence of 
the coexistence of the two gases, 
there is a linear temperature component in the electronic 
specific heat, which comes from (at least) the normal part of the system, 
while the quadratic temperature term is a contribution of the bosonic superconducting 
counterpart. 
In some articles \cite{Indios} it has been suggested that the linear component 
comes exclusively from the presence of holes, however, in this version of our model 
we can not discriminate between holes and electrons. 
And last, we are able to state that the contribution of the electronic 
specific heat to the total is around 30$\%$, and not only 1 - 2$\%$ 
as has been repeatedly suggested \cite{Loram93,Meingast09}.

\section{\label{Lattice}Lattice specific heat}

In this section we calculate the specific heat due to the lattice $C_l$, which 
represents the major contribution for any solid. To do so, we 
describe and use two different formalisms: a simple Debye model and a 
phenomenological procedure where we take the phonon spectrum from inelastic 
neutron scattering experiments, and compare both results. 
After this, we proceed to add the resulting curve to our previous electronic 
specific heat results.

The total internal energy of a crystal is in general obtained by taking the normal 
vibration mode number $G(\omega) d\omega$ which lie in an interval $\omega$, 
$\omega + d\omega$ with a frecuency $\omega$, where $G(\omega)$ is the phonon 
density of states (PDOS). We take the energy $\hbar\omega$ of each mode, so 
\cite{Kittel}
\begin{equation}
U={\int }\frac{\hbar \omega G(\omega )d\omega }{\left( \exp [\hbar \omega /
{k_{B}T}]-1\right) },  \label{IntEnerLatt}
\end{equation}%
where we have the assumptions that the crystal is large enough so the sums can 
be substituted by integrals as usual. The specific heat for the lattice is
\begin{equation}
C_{l}=k_{B}{\int }\frac{(\hbar \omega /{k_{B}T)}^{2}\exp [\hbar \omega /
{k_{B}T}]G(\omega )d\omega }{\left( \exp [\hbar \omega /{k_{B}T}]-1\right)
^{2}},  \label{SpecHeatLatt}
\end{equation}
which is the expression from which the two methods we use for our analyses are 
derived.

\subsection{Debye model}

In the Debye model the solid is considered monoatomic and isotropic.
The phonon density of states is  $G(\omega )=3k^{2}/2\pi ^{2}\mathsf{v_0} 
= 3\omega ^{2}/2\pi^{2}\mathsf{v_0}^{3}$, where $\mathsf{v_0}$ is the sound 
velocity, which has been supposed to be the same in any direction (transversal 
or longitudinal). Introducing this PDOS in (\ref{SpecHeatLatt}), and after 
some algebra we have the Debye expression for the lattice specific heat 
\begin{equation}
\frac{C_{l}}{N_lk_{B}}=9s\left( \frac{T}{\Theta _{D}}\right) ^{3} 
{\int_{0}^{\Theta _{D}/T}}\frac{\chi^{4}\exp [\chi] d\chi}{\left( 
\exp[\chi]-1\right) ^{2}},  \label{Cvdebye}
\end{equation}
where $\Theta _{D} \equiv \hbar \omega _{D}/k_{B}$ is the Debye temperature 
characteristic of every solid, $\chi \equiv \hbar \omega_D /{k_{B}T}$, $N_l$ the 
number of unit cells in the solid and 
$s=13$ is the number of atoms per unit cell. 
We must remark that in this calculations we
are assuming that the Debye temperature of YBa$_{2}$Cu$_{3}$O$_{7-x}$ is a
constant over the complete interval of temperatures considered.

\begin{figure}[htb]
\begin{center}
\centerline{\epsfig{file=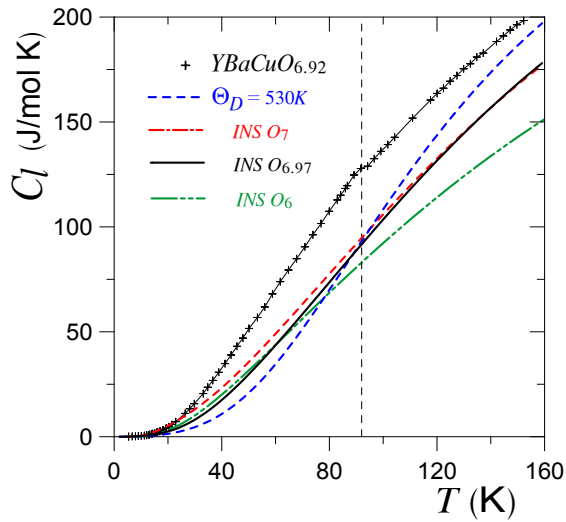,height=2.8in,width=3.0in}}
\end{center}
\par
\vspace{-1.0cm}
\caption{ Lattice specific heat for YBa$_{2}$Cu$_{3}$O$_{7-{x}}$ 
using the Debye model for  $\Theta _{D}$ = 530 K and inelastic 
neutron scattering (INS) for three doping ${x}$. 
For comparison we add the the total specific heat experimental data (crosses) reported for  
YBa$_{2}$Cu$_{3}$O$_{6.92}$ by \cite{Bessergenev}.}
\label{fig:CvDebyeandNIS}
\end{figure}

When one uses the Debye model for lattice calculations, it is implicit that 
one is taking the harmonic oscillator approximation, so in an intent to go
beyond, anharmonic terms such as cubic and fourth order in $T$ should be 
considered. However, several authors report that the anharmonicity does not
exceed the $10$ to $15$\% of the electronic specific heat component in the 
$0$ to $300$ K interval \cite{Bessergenev,Naumov}, so it is usually 
ignored. 
Although Debye's formalism reproduces remarkably well the $T^3$ behavior at 
$T < 5$ K for cuprates, and is successful for monoatomical 
superconductors, such as Al \cite{ATari}, it is not longer good for cuprates 
above 5 K, as we show in Fig. \ref{fig:CvDebyeandNIS}, where we present the 
lattice specific heat as a function temperature together with the experimental 
specific heat reported by Bessergenev \textit{et al} for YBa$_{2}$Cu$_{3}$O$_{6.92}$ 
\cite{Bessergenev}. Even though the most commonly associated Debye 
temperature for this cuprate is $\Theta _{D}$ = 420 K, from now on we use the 
value $\Theta _{D}$ = 530 K in the knowledge that it provides better results, as 
will be seen later. 
The vertical dashed line in the figure indicates the critical temperature 
$T_c$ = 92.2 K.

\subsection{Inelastic neutron scattering}

An alternative way to use series and approximations is to appeal to
the experimental phonon density of states (PDOS) reported by inelastic 
neutron scattering (INS) for YBa$_{2}$Cu$_{3}$O$_{7-x}$ with several doping 
${x}$ \cite{Renker88}. In their experiments, the authors use a non 
superconducting sample as a reference, such as YBa$_{2}$Cu$_{3}$O$_{6}$, 
which is also reported. 

We extract the PDOS data for YBa$_{2}$Cu$_{3}$O$_{6.97}$, YBa$_{2}$Cu$_{3}$O$_{7}$ 
and YBa$_{2}$Cu$_{3}$O$_{6}$ from the curves given in \cite{Renker88}, 
introduce each one in Eq. (\ref{SpecHeatLatt}) 
to perform the integrals numerically observing the correct handling of the
units, and include the curves in Fig. \ref{fig:CvDebyeandNIS}. We are assuming 
that the use of this approach already takes into account the anharmonic terms, 
at least up to the temperature interval considered.

\begin{figure}[htb]
\begin{center}
\centerline{\epsfig{file=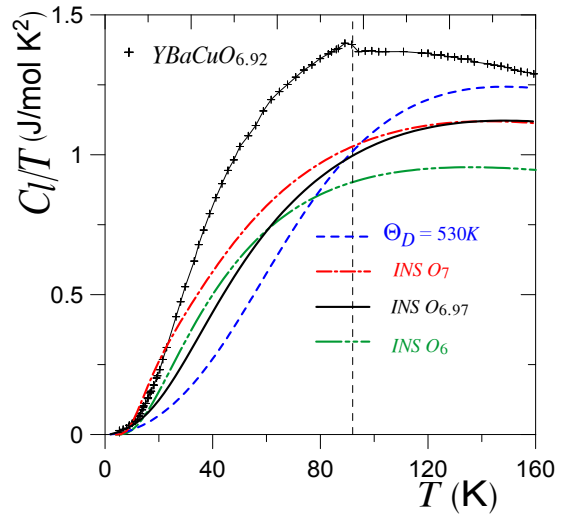,height=2.8in,width=3.0in}}
\end{center}
\par
\vspace{-1.0cm}
\caption{Specific heat over temperature of the phonons using the Debye model
and Inelastic Neutron Scattering for the same values as in Fig. 
\ref{fig:CvDebyeandNIS}.}
\label{fig:CvsTDebyeandNIS}
\end{figure}

To magnify the difference among the lattice specific heat curves  for 
different oxygen concentrations, we show in Fig. \ref{fig:CvsTDebyeandNIS} 
the lattice specific heat over temperature for the same values used in Fig. 
\ref{fig:CvDebyeandNIS}, keeping in mind that this type of curve is the most 
used one for reporting electronic specific heat.  
It can bee seen that the differences between experimental results and
theoretical curves are more noticeable in this form. At this point we would 
like to analize three features: the difference between the total experimental 
specific heat for YBa$_{2}$Cu$_{3}$O$_{6.92}$ and the lattice specific heat computed 
using the PDOS for $7-x = 6.97$ at the transition point is around 30$\%$ in 
this graphic; however, the difference in the lattice specific heat between two 
adjacent superconducting dopages at the same point is quite small; and 
last, the lattice specific heat for $7-x=6$ is 10$\%$ lower at $T_c$ than that 
for the superconducting counterparts. On these bases we conjecture that using 
the YBa$_{2}$Cu$_{3}$O$_{6}$ compound as a reference for the specific heat of 
the lattice can be considered as a rough approximation. 

Above $T_c$ the lattice specific heat we obtain is still valid, but, 
as we mentioned before, our curve for the electronic specific heat is 
not strictly accurate, but only a guide.

By plotting the lattice specific heat $C_l/T$ vs $T^2$, we are  
able to reproduce the $T^3$ term observed in some experiments 
\cite{WangPhysRevB2001}. As we said above, this is true for the 
Debye model, as expected, but using the INS phonon density of states 
we find a $\beta$$T^3$ behavior for $T <$ 5 K with $\beta$ = 0.683 mJ/mol 
K$^{4}$, compared to 0.305 and 0.392 reported in \cite{WangPhysRevB2001} 
and \cite{Moler97} respectively. 
Notice that our value of $\beta$ lies within the uncertainty of the INS 
experiment, and that unfortunately, no further experiments on PDOS for 
cuprates have been reported.

\section{\label{Sum} Specific heat of YB\lowercase{a}$_{2}$C\lowercase{u}$_{3}$O$_{6.92}$}

\begin{figure}[htb]
\begin{center}
\centerline{\epsfig{file=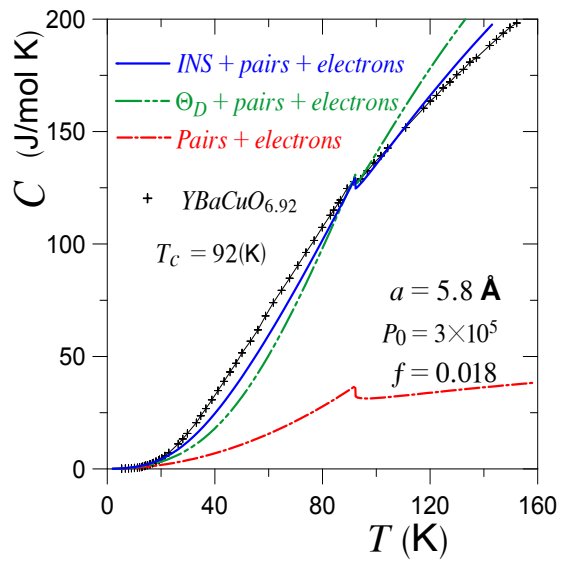,height=3.in,width=3.0in}}
\end{center}
\vspace{-1.0cm}
\caption{Total specific heat for YBa$_{2}$Cu$_{3}$O$_{6.92}$.}
\label{fig:Cvlineal}
\end{figure}

\begin{figure}[htb]
\begin{center}
\centerline{\epsfig{file=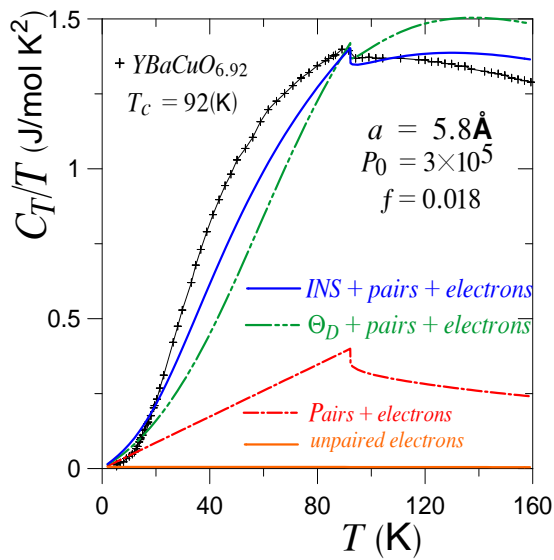,height=3.in,width=3.0in}}
\end{center}
\par
\vspace{-1.0cm}
\caption{Total specific heat over temperature for YBa$_{2}$Cu$_{3}$O$_{6.92}$. }
\label{fig:CvsTLineal}
\end{figure}

We take the electronic specific heat we calculated for paired and unpaired 
electrons with the parameters $P_{0} = 3 \times 10^5$ and $f = 0.018$, 
and add them to the lattice specific heat from the Debye model with 
$\Theta_D$ = 530 K and from the INS spectra for YBa$_{2}$Cu$_{3}$O$_{6.97}$, 
which we will be using in our calculations from now, 
and plot them in Figs. \ref{fig:Cvlineal} and  \ref{fig:CvsTLineal}. Together 
with the total specific heat we plot the electronic specific heat (normal plus 
superconducting) for the purpose of emphasizing the size of its contribution. 

From these figures we may observe that the total height at the transition point 
for both, specific heat and specific heat over temperature, are reproduced 
by adding the three components we analized. However, we observe a minor difference 
between the experimental shape of the curve and ours below $T_c$, more remarkable 
in Fig. \ref{fig:CvsTLineal}. This difference becomes more notorious 
around 40 K, where we believe that the contribution of the lattice $C_l$ 
needs a better analysis, supported by a more accurate experiment.

\section{\label{conclusions}Conclusions}

While most authors take the experimental curves of the total specific heat
and subtract components, we are able to qualitatively construct the total
specific heat for the YBa$_{2}$Cu$_{3}$O$_{7-x}$ cuprates from a simple first
principles model: the Boson-Fermion theory of superconductivity with the layered 
structure of HTSC. The model consists in taking the Cooper pairs as a boson gas 
coexisting with a unpaired electrons (or holes) Fermi gas, both gases under the 
confinement of the layered structure modeled by a Kronig-Penney potential 
in the Dirac comb limit in the perpendicular  direction to the CuO$_{2}$. Although no interactions among 
bosons and no internal and/or external magnetic fields are considered, the model 
reproduces qualitatively well the experimental curves of the total specific heat. 

A direct result is the critical temperature which depends on the anisotropy of 
the material, introduced through the planes impenetrability $P_0$ and the separation $a$ 
between them. Since we assumed that not all the pairable fermions are paired, this critical temperature also depends on 
the fraction of fermions that formed Cooper pairs. 
The total specific heat is calculated by including the specific heat coming from 
the bosons (superconducting electronic specific heat), unpaired fermions (normal 
electronic specific heat) and the lattice. The resulting curve is compared to 
that of  YBa$_{2}$Cu$_{3}$O$_{6.92}$, which at $T_c$ = 92.2 K has 
a jump $\Delta C/T_c$ = 20 mJ/mol K$^2$ and is characterized by a linear dependence on 
temperature $\gamma_e T$ and a quadratic one $\alpha T^2$.
These last two features are reproducible with our model by fixing the parameters 
$P_0$ and $f$ with the known values of $T_c$ and $\Delta C/T_c$ at $T_c$.
The values we get for the constants $\gamma_e =$ 5.2 mJ/mol K$^2$ and 
$\alpha$ = 4.3 mJ/mol K$^3$ are of the order of the experimentally reported ones. We also show their correspondence with the normal 
electronic specific heat coming from the unpaired fermions, and the 
superconducting term from the paired fermions, respectively. At the same time, 
these two results make plausible the assumption that not all pairable fermions in 
the Fermi sea were paired, even at temperatures near zero, and that the jump 
is a consequence of the condensation of the pairs. 
We also show that a simple Debye model for the lattice specific heat fails, 
as expected, but considering the results from INS experiments gives a better, 
albeit not perfect, approximation. A closer shape of our lattice specific heat 
to the experimental one should be obtained using data from a more accurate 
INS experiment. 
It can also be seen that the lattice specific heat shows the same temperature 
cubic behavior for $T <$ 5 K as experiments show \cite{WangPhysRevB2001}.

Another important result of our analysis is that  the 
electronic specific heat (normal plus superconducting ) has a contribution of $30 - 40\%$ of the total at the 
transition temperature, and not only the $1 - 2\%$ most authors consider, which 
is of the order of the specific heat of the unpaired fermions alone. We suggest that  
when they subtract what they consider the lattice specific heat from a 
sample, either constructed by fits or using a non-superonducting reference material, they 
might be taking away a representative part of the total electronic specific heat.

Finally, we indirectly confirm that the upturn in the total specific heat at very low temperature is not a result of paired or unpaired fermions in the absence of a internal or external magnetic field. However, including magnetic terms should be a starting point for future study.


We acknowledge the partial support from grants CONACyT 104917 and PAPIIT IN-111070 and IN-105011.

\end{document}